\documentclass[10pt, a4paper]{article}

\usepackage{url}
\usepackage{authblk}
\usepackage{graphicx}
\usepackage{pdfpages}

\title{The Sun and its educational spectrum}

\author[1,2]{Alejandro C\'ardenas-Avenda\~no}
\affil[1]{Programa de Matem\'atica, Fundaci\'on Universitaria Konrad Lorenz, 110231 Bogot\'a, Colombia}
\affil[2]{eXtreme Gravity Institute, Department of Physics, Montana State University, 59717 Bozeman MT, U.S.A.}
\author[3]{Santiago Vargas Dom\'inguez}
\affil[3]{Observatorio Astron\'omico Nacional, Universidad Nacional de Colombia, Bogot\'a, Colombia}
\author[4]{Freddy Moreno-C\'ardenas}
\affil[4]{Centro de Estudios Astrof\'isicos, Colegio Gimnasio Campestre, Bogot\'a, Colombia}
\author[3]{Benjam\'in Calvo-Mozo}

\date{\today}

\begin{document}

\maketitle

\begin{abstract}

The aim of this paper is to encourage science educators and outreach groups to look appropriately at the Sun and consider it as an extraordinary pedagogical tool to teach science at all education stages, what we call here as the solar educational spectrum, i.e., from K-12 to higher education, to develop informal educational projects that may lead to reach more complex material and to enlarge the experience at each stage. We review the main aspects of the Sun as an appetizer of the endless source of ideas to perform informal educational projects outside of a structured curriculum. We end up our discussion by sharing our experience across the educational spectrum in Colombia and how we used it as a development instrument.

\textbf{Keywords:} Public engagement with science and technology; Science communication in the developing world; research-link; Informal education; The Sun.

\end{abstract}

\section{The Sun and its active habits }

The Sun has undeniably ruled our lives since the beginning of human
history, or even previously, from the very birth of the Earth. During
the youngest stage of our star, about 4.6. billion years ago, before
planets in the solar system were even formed, all the matter evolved
around the Sun as a huge hot disk. This protoplanetary disk was in
constant cooling and the planets emerged in areas where matter got
more concentrated due to gravitational attraction. Millions of years
later the conditions improved to favor the development
of life. In the same way as its origin, the fate of our planet is strongly linked
to the Sun which will be responsible also for its annihilation, but we
will come back to the final stage of our blue planet at the end of
this short review.

Let's go back to the past once again, but now to Italy and the epoch
of the great Galileo Galilei in the early XVII century \cite{Vliegenthart1965}.
Soon after the invention of the telescope by the German optician Hans
Lippershey (although there are still some other names that are claimed
to be the inventors) \cite{Helden2010}, Galileo designed his own
telescope and pointed it to the sky. Initially, the potential of the
telescope was hard to imagine, but eventually this invention changed
our conception of the universe.

\begin{figure}
\begin{centering}
\includegraphics[scale=0.65]{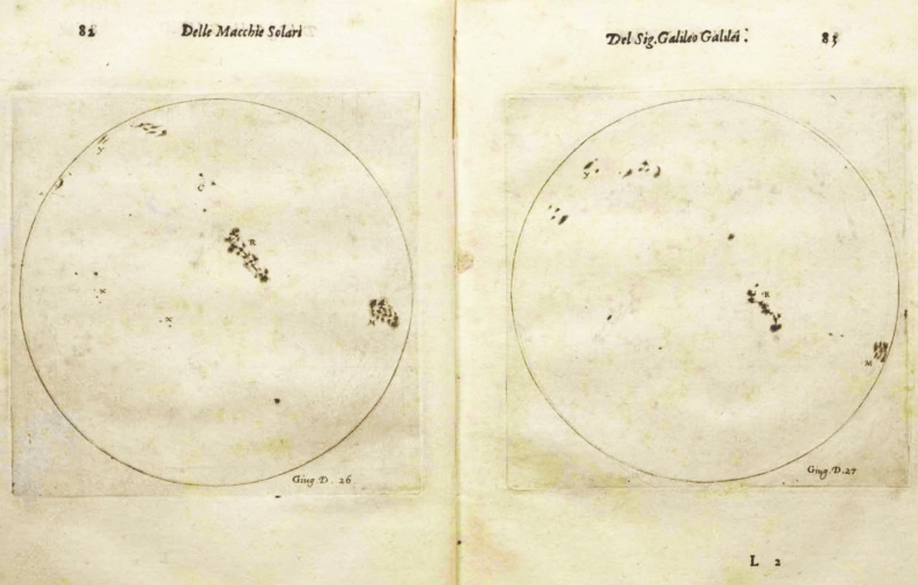}
\par\end{centering}
\caption{\label{fig:Drawings-of-the}Drawings of the solar disk displaying
sunspots made by Galileo Galilei at the beginning of the XVII century,
published in 1613 in \textquotedblleft Istoria e Dimostrazioni Intorno
Alle Macchie Solari e Loro Accidenti\dots \textquotedblright  (History
and Demonstrations Concerning Sunspots and their Properties) \cite{Galilei1613}.
Galileo discovered that these dark areas where located on the Sun
although could not figure out precisely what was causing them.}

\end{figure}

Galileo pursued, among others, some of the first solar observations
with the new optical tool, by projecting the image of the Sun through
the telescope, and discovered something wonderful and unexpected:
dark scars on the surface of the star, see Fig.~(\ref{fig:Drawings-of-the}).
He was looking at what we now call sunspots, which in fact had been
observed by naked eye (with very little detail) well before Galileo
\cite{Blofeld1958}, but were thought to be planets revolving around
the Sun in very-close orbits and not actual structures placed on the
Sun. Galileo's observation of sunspots had major implications for
the thinking of the time, not only on the scientific ones but also
on religious conceptions, since it was showing imperfections of heaven,
and in particular of the Sun, that was especially considered divine
and therefore perfect. Galileo\textquoteright s detailed drawings
showed the time evolution of sunspots and evidenced the rotation of
the Sun \cite{Benz2017}.

Four centuries after Galileo's seminal works, sunspots are still a
target of study and are highly correlated with the solar activity,
which is responsible, in turn, for the conditions of the interplanetary
medium, in what is commonly referred to as space weather. It is known
that the Sun is composed of plasma at very high temperature, of about
5,700 Kelvin on the surface and up to 15 million Kelvin in the interior,
and that sunspots are a visible manifestation of the magnetism of
the star. 

\subsection{The explosive Sun }

The Sun is like a huge magnet with its north and south poles forming
a global dipolar magnetic configuration, although it also displays
local magnetism in all its surface. Every twenty-two years something unexpected happens: the solar magnetic poles are reversed and it is called as the magnetic cycle. Half of this time interval is a solar cycle, i.e., eleven years, and it was discovered in 1843 by Samuel
Heinrich Schwabe, who after 17 years of solar observations noticed
a periodic variation in the average number of sunspots per year \cite{Schwabe1844}.
During the solar cycle, variations in the solar activity and the amount
of radiation reaching our planet are related, with extreme periods
of low solar activity -solar minimums- and high activity -solar maximums-.
When the solar minimum occurs, sunspots disappear almost entirely
from the Sun's surface while in peak periods there are plenty of them.
Currently, the Sun is in its 24th cycle, counting from the first cycle
registered from 1755 to 1766. 

Magnetic fields are formed inside the star, and emerge through the
solar surface (photosphere), interacting with the glowing plasma.
The magnetic field lines are constantly evolving, largely because
the Sun rotates faster at the poles than at equator. This differential
rotation stretches and twists the field lines in the solar interior,
and eventually generates instabilities that end up on the surface,
forming the sunspots \cite{Benz2017}. 

When the Sun's magnetic field changes its configuration along the
solar cycle, energy can be released rapidly and violently once magnetic
field lines emerge and interact with the plasma in the solar atmosphere
in events known as coronal mass ejections. During these bursts the
Sun throws away thousands of tons of plasma and charged particles
into the interplanetary medium. If the blast is directed towards Earth,
the solar storm could strike it and generate what is called a geomagnetic
storm. Without the Earth's magnetic field, a natural shield called
the magnetosphere, we would be seriously affected by radiation from
the Sun. Nevertheless, some of these charged particles penetrate the atmosphere
in polar regions, which are the most sensitive areas of the Earth's
shield and where magnetic field lines converge, producing amazing
phenomena such as aurorae. 

\begin{figure}
\begin{centering}
\includegraphics[scale=0.50]{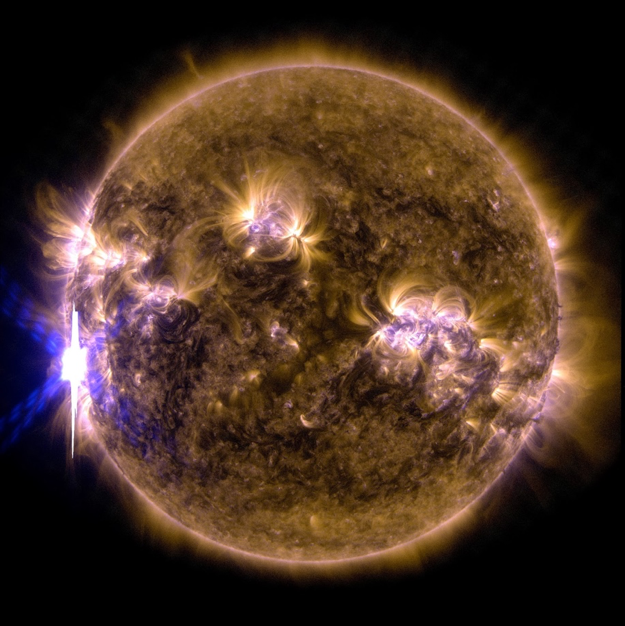}
\includegraphics[scale=0.08]{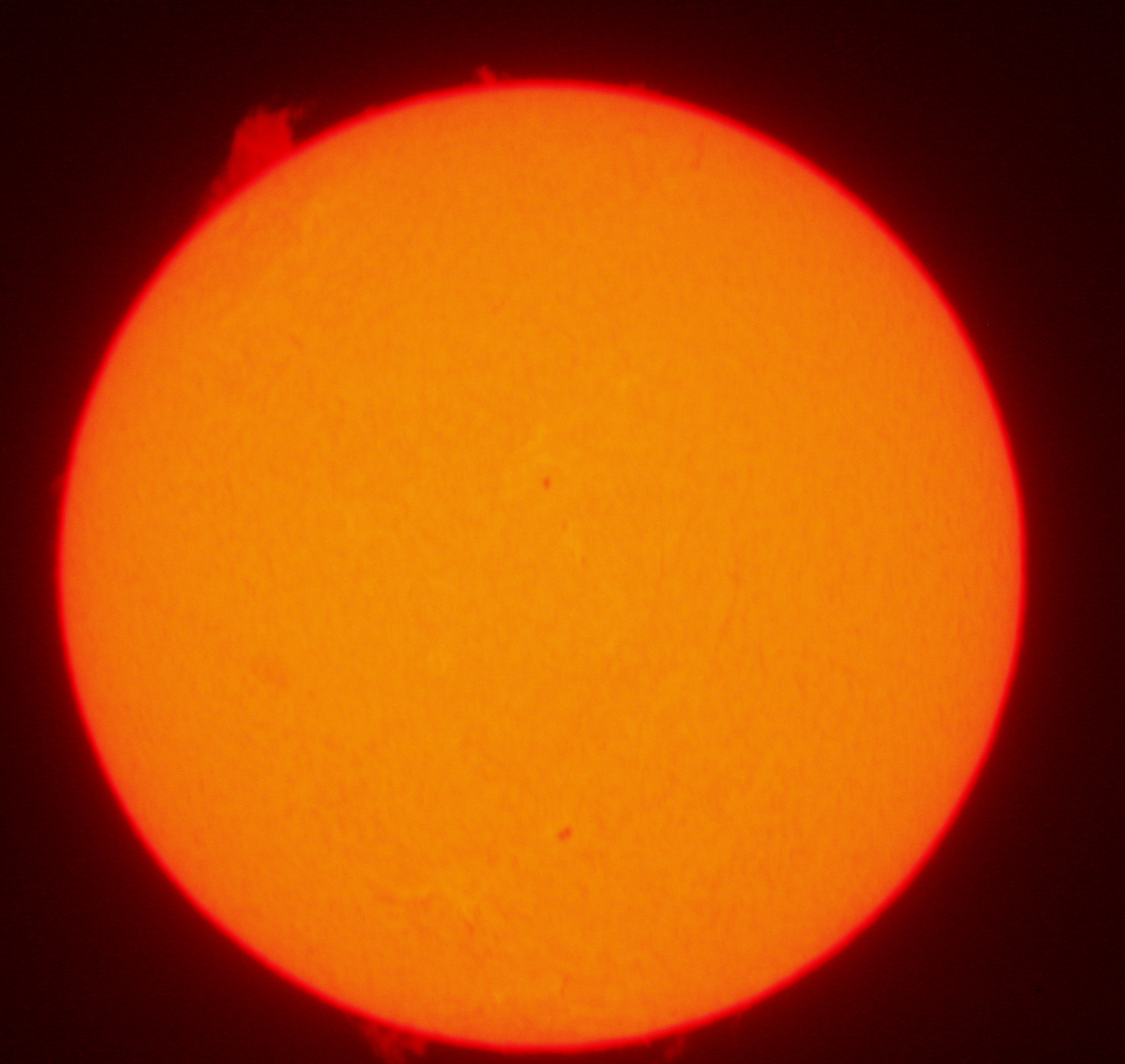}
\par\end{centering}
\caption{\label{fig:Image-of-the}Image of the Sun taken with the Solar Dynamics
Observatory (SDO) \cite{Pesnell2012} on February 24, 2014. The region
on the left displays an X-class solar flare emitting large amount
of energy (Left). Prominence taken by Freddy Moreno-C\'ardenas on April 22, 2015 at Observatorio Julio Garavito, Gimnasio Campestre, Bogotá, Colombia (Right)}
\end{figure}

Other phenomena can generate gigantic solar flares that could release
extraordinary amounts of radiation to our planet, see left panel in Fig.~(\ref{fig:Image-of-the}).
In March 1989, about six million people in Canada and in the US lost
power after a huge explosion on the Sun that affected a hydroelectric
plant in Quebec, affecting a large amount of electrical transformers
\cite{Allen1989}. 

The largest solar storm ever recorded occurred in 1859 and is known
as the perfect storm or the Carrington Event, honoring the name of
a British scientist who detected it. At that time the consequences
were in serious damage of the North America telegraphic system, that
had 15 years of being invented, and observation of aurorae at very
low latitudes such as Florida and even Colombia \cite{MorenoCardenas2016}. 

\subsection{Understanding the solar behavior }

One of the greatest scientific challenges of our time is to understand
and accurately predict the solar cycle and the effect of changes in
solar activity on the climate of the planet, as there are tests that
suggest a nontrivial dependence \cite{Lockwood2009}. The most famous
documentation constitutes a period of 70 years (1645-1715) in which,
according to records, there were virtually no spots on the solar surface,
period often called Maunder Minimum. Another period of scarcity of
sunspots has been named Dalton Minimum (1795-1825). These intervals
agree with the dominant part of a period of intense cold in Europe
known as the Little Ice Age (1450-1850), but the actual connection
is still under debate \cite{Mann2009}.

Regarding abnormal solar cycles, the previous one (number 23) which
began in 1996, was also quite unusual and spanned a couple of years
more than average. Starting in 2009 the Sun awakened from its slumber
and sunspots began to appear again on its surface, which marked the
beginning of cycle 24. However, opposite to most of the predicted scenarios \cite{Svalgaard2005}, the current solar magnetic activity cycle has been quite weak and the number of sunspots retained below 100, unlike other cycles which registered more than 250, e.g., the current cycle 24 peaked around the year 2014, with nearly half the amplitude of the previous one. 

It has been said that magnetism is responsible for the majority of
events taking place in the Sun and ultimately for space weather, i.e.,
the set of phenomena and interactions occurring in the interplanetary
medium are regulated primarily by solar activity. Bearing in mind
that the Sun is almost 99\% of the mass of the entire solar system
and therefore their crucial role on impacting our cosmic neighborhood
and directly on planet Earth. For these reasons, scientists continue
to explore the complex solar activity with the aim to understand and
predict the behavior of the Sun. Currently, we have sophisticated
telescopes and instruments, many of which observe the Sun minute by
minute from space, through satellites orbiting the Earth, giving us
spectacular images that demonstrate its intricate structure and register
lots of explosive phenomena. Recently, a NASA mission launched in
2010 with the Solar Dynamics Observatory (SDO) \cite{Pesnell2012},
began acquiring high-quality solar images with 10 times better resolution
than that of an HDTV, taking up to 1.5 terabytes of data each day,
equivalent to more than half a million songs in an mp3 player. 

The Sun has been the main source of energy for maintaining life on
Earth; it has gone from being a source of worship to an object of
study for human beings. There will come a day when the Sun exhausts
all its fuel, expanding and becoming a red giant star, whose size
is so large that it will possibly swallow up the Earth. The final
destination of the Sun is to become a white dwarf, and eventually
a black dwarf when cooled completely. There are still about five billion
years for this to happen, but for now we must prepare for each solar
maximum and to protect our technology, on which our modern civilization
depends. We should not be catastrophic, but always keep in mind
that a perfect solar storm, as the one that happened in 1859, could
now have a remarkable impact on satellite technology, communications,
power grids and others, causing millions of dollars in losses. 

\section{Solar educational spectrum, i.e., from K-12 to higher education, in Colombia}

Colombia has experienced a scientific, technological and cultural
transformation in recent years, showing a considerable growth in these
areas. Being in the midst of such important changes, it is critical
to open spaces for the recognition of the importance of science and
its public perception, and ultimately for the formation of future
science communicators capable of impregnating the society with passion,
admiration and love for what science represents and what it can offer.
In this section we will describe the formal and informal educational experiences that we have had in a high school program, extracurricular science clubs and at the University, in order to provide learners with the tools they need to  reach more complex material and to enlarge their experience at each stage. 

\subsection{Observing the Sun at school to motivate scientific skills }

\begin{figure}
\begin{centering}
\includegraphics[scale=0.65]{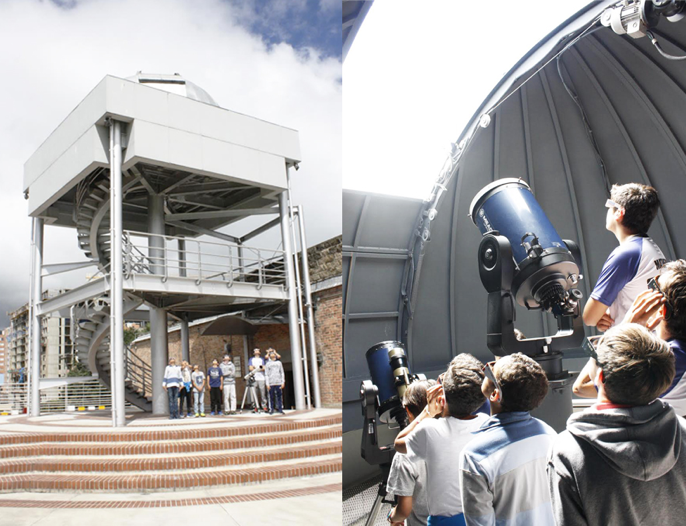}
\par\end{centering}
\caption{\label{fig:Group-of-students}Group of students at Julio Garavito
Armero Observatory founded in the year 2000 at the high school Gimnasio
Campestre in Bogot\'a, Colombia.}
\end{figure}

Gimnasio Campestre, a school at Bogota, Colombia, has developed since 1997
a plan for teaching basic astronomy. Its astronomical observatory,
named Julio Garavito Armero honoring the name of a prominent Colombian
astronomer from the twentieth century, was built in 2000, see Fig.~(\ref{fig:Group-of-students}). The observatory develops several activities
and classes to involve children of different ages into science. For
the youngest students, the curriculum is designed on learning the
basics about orientation, constellations and observations of the Sun,
Moon and planets. Then, starting from fourth grade, the classes cover
three main subjects: a.) origin of matter and the universe, b.) the
light and the Sun and c.) gravity and asteroid impacts. In sixth grade
the school offers a class on: a.) stars and the Sun, b.) Earth (rock
and water, planetary geology) and c.) planetary atmospheres. After
taking these classes, students from seventh to eleventh grade, the
last one in these educational system, have the option to deepen into
the topics covered and do research on a curricular unit called \textquotedblleft J\'ovenes
Investigadores\textquotedblright{} (young researchers).

Teachers have found that Sun is an interesting topic in which students can get involved  every year. In this academic environment, students learn about multiple characteristics and
phenomena of our star and also get trained in astrophotography. In
particular, the observatory keeps a record of solar observations since
2001. An important milestone in this project was the acquisition of
a H-alpha solar telescope, which allowed to acquire images of prominences,
filaments and solar flares. Every two years the group of teachers and students involved in registering  the evolution of the solar cycle publishes an article on this subject in the school's research journal named
\textquotedblleft El Astrolabio\textquotedblright \footnote{http://www.revistaelastrolabio.com}
and some of the most interesting images are sent to the Spaceweather
website\footnote{http://spaceweather.com }, see right panel in Fig.~(\ref{fig:Image-of-the}). The results of the developed projects have been used to publish six articles at the school's research
journal. Furthermore, several projects have enabled students and teachers
to participate and present their research at conferences and national
astronomy meetings, and even publish in international journals, such as the one where it was discovered a report about an aurorae borealis seen in Colombia in 1859 \cite{MorenoCardenas2016}. 

The observation of the Sun as basis of a school research project,
has been replicated in several schools of Bogot\'a. Currently, Gimnasio
Campestre's Observatory mentors two public schools, fostering the
development of scientific knowledge through the involvement of students
and teachers in lively projects. 

\subsection{Studying the Sun before going to college: Seedbed projects of research }

The Young Talent Program of Mathematics, Science and Technology takes
place each year, since 2012, at Fundaci\'on Universitaria Konrad Lorenz in Bogot\'a,
Colombia. This program is a non-profit project that focuses on high-school
children from Bogot\'a and cities around, including some from disadvantaged
communities, designed to inspire young people and bring them into
science, with astronomy as the vehicle.

Every year the group is formed by 30 participants, selected by an
admission exam which measures the ability of solving problems without
previous particular knowledge, from an application pool of over 200
young people. The participants are supported during one year by two
instructors whose propose is answering questions and to propose a
main objective at the beginning of the program, with no lectures at
all.

During 2014 the team\textquoteright s general objective was to learn
scientific computing techniques. The chosen programming language was
Python, in order to exploit its expressive power with simple and compact
syntax, and the third-party open-source libraries such as Numpy \cite{Walt2011},
Matplotlib \cite{Hunter2007} and SciPy \cite{Jones2014} to encourage the use of scientific scripting language. 

\begin{figure}
\begin{centering}
\includegraphics[scale=0.85]{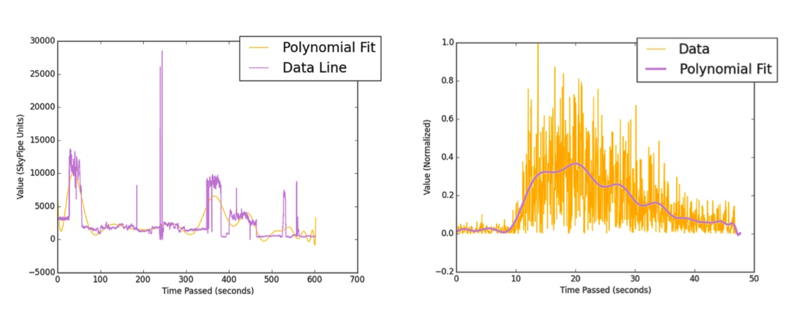}
\par\end{centering}
\caption{\label{fig:An-example-of}An example of scientific analysis performed
by a group of the young participants from data taken on September
6, 2014 (14:10-14:20 UT) aimed to implement a simple polynomial fit
by least squares using Numpy \cite{Walt2011} functions.}
\end{figure}

In order to develop scientific computing techniques, the proposed
objective was to measure the solar activity via radio waves with NASA's
Radio Jove project antennas \cite{Lashley2010,Higgins2014}, without using the main software provided by NASA to record, store and visualize the data. Fig.~(\ref{fig:An-example-of}) shows an example
of the data analysis performed by the participants. 

In addition to the data collected by the students, the Radio JOVE
Data Archive\footnote{The Radio JOVE Data Archive, http://radiojove.org/archive.html http://radiojove.gsfc.nasa.gov/data\_analysis/.}
was used in order to gather more information and explore bigger data
sets. In some cases, it was possible to make a rigorous analysis and
check if the registered phenomena were global, at least reported somewhere
else, or corresponded to spurious signals. With all the data collection
and analysis, the participants were able to recognize a relationship
between sunspot numbers and 20 MHz solar burst counts.

\begin{figure}
\begin{centering}
\includegraphics[scale=0.65]{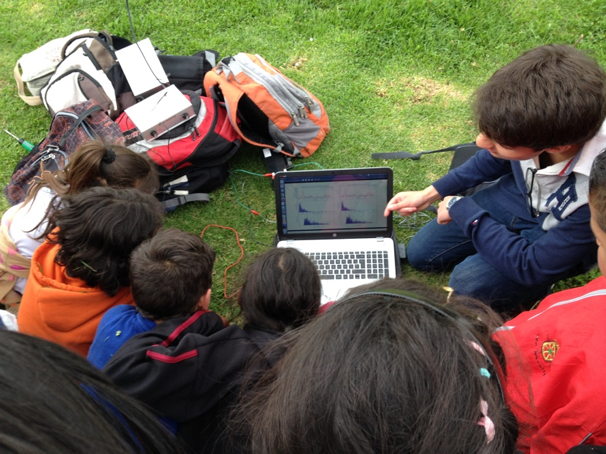}
\par\end{centering}
\caption{\label{fig:A-group-of}A group of students showing their collected
data and performing in open spaces where participants are encouraged
to show to the public, specially to young children, what they were
doing.}
\end{figure}

The results were presented in Regional STEM Fairs open to the general
public and all the data collection was done in public parks in order
to engage young children and the general public, see Fig.~(\ref{fig:A-group-of}).
During the data collection campaigns, the participants make shifts
between collecting data and explaining the project to passers-by. 

We also would like to mention that two participants of this project
performed best at the Colombian Olympiad in Astronomy and then were
part of the National Team for the International Olympiad in Astronomy
and Astrophysics (IOAA) and the Latin-American Olympiad of Astronomy and
Astronautics (LAOAA). 

\subsection{Solar physics as a research option at university }

\begin{figure}
\begin{centering}
\includegraphics[scale=0.75]{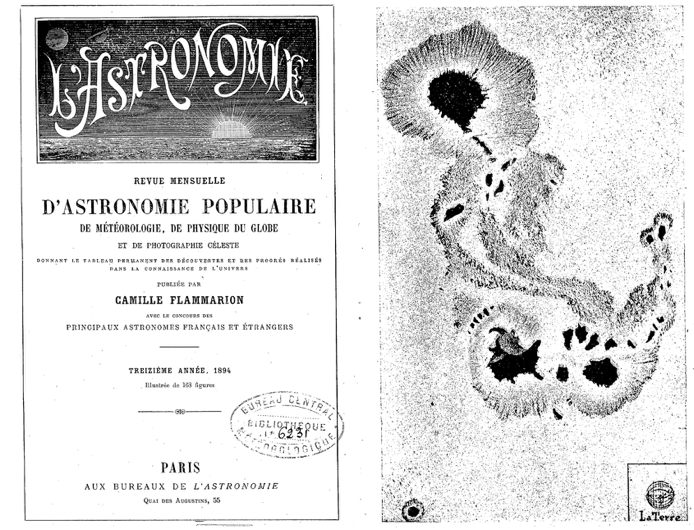}
\par\end{centering}
\caption{\label{fig:Journal-of-the}Journal of the French Society of Astronomy
(left) in which Jos\'e Mar\'ia Gonz\'alez Benito published a drawing of
the large solar active region observed in August 1893 (right) \cite{Flammarion1896}.}
\end{figure}

The study of the Sun in Colombia dates back to the observations made
by Jos\'e Mar\'ia Gonz\'alez Benito in the 19th century, when he was the
director of the National Astronomical Observatory (the oldest astronomical
observatory in America founded in 1803) and member of the French Society
of Astronomy. Gonz\'alez published a drawing of the large sunspot of
August 1893 in the journal of the society, see Fig.~(\ref{fig:Journal-of-the}).
Nowadays, Colombia has consolidated astronomy research options in
different universities across the country. In this section we will
only refer to the experience acquired at the National Astronomical
Observatory (that belongs to Universidad Nacional de Colombia). 

From its begin in the year 2011, the Group of Solar Astrophysics
(GoSA) has convened undergraduate and graduate students. Starting
from a seed of two students enrolled in the undergraduate program
in Physics that among their interest in multiple areas of astronomy
got strongly motivated for solar physics, and with the help of one
of the astronomy professors at the Observatory, the GoSA got involved
in the analysis of solar data.

In 2012 with the organization of an International Summer School entitled
"Solar Astrophysics: Modern trends and techniques"
held at the National Astronomical Observatory, a great number of undergraduate
students in Physics joined the GoSA and decided to pursue their research
project (a requirement to get the university degree)
in solar physics. In order to do so, students were trained in programming
languages of common use in the area, e.g. Interactive Data Language
(IDL) \cite{Landsman1993} and started analyzing data from cutting-edge
satellite solar telescopes. Working with time series of images, they
explored topics ranging from tracking of solar spicules to the study
of hard X-ray in solar flaring events.

GoSA members are currently engaging with the challenges and opportunities
of doing state-of-the-art research and are involved in international
collaborations. A number of 7 master students, and 12 undergraduate
students are part of the research project entitled "Magnetic
field in the solar atmosphere" that comprises individual
research topics dealing the with the analysis of ground-based telescopes,
such as the Solar Swedish Tower, SST \cite{Scharmer2003}, and space
facilities like SOHO \cite{Domingo1995}, RHESSI \cite{Lin2002}, SDO \cite{Pesnell2012} and Hinode \cite{Kosugi2007}. Currently
the group is developing routines for data analysis in Python as part
of the Sunpy collaboration \cite{Sunpy2015}.

In 2015 the university included a course on solar physics in its official
academic program with the name "Foundations of Solar
Astrophysics", intended for master students but also
with the option to be taken by undergraduates. An average of 15 students
are taking the course every semester, and a high percentage of them
got motivated and joined the research group.

More recently, the GoSA organized the International Astronomical Union
Symposium 327 (IAUS327) entitled "Fine Structure and
Dynamics of the Solar Atmosphere" that was held in Cartagena
de Indias, Colombia, 9-13 October 2016. This event consolidated the
group and was a foremost opportunity to increase the visibility of
the members and their research works among the international community,
therefore promoting new collaborations.

\begin{figure}
\begin{centering}
\includegraphics[scale=0.65]{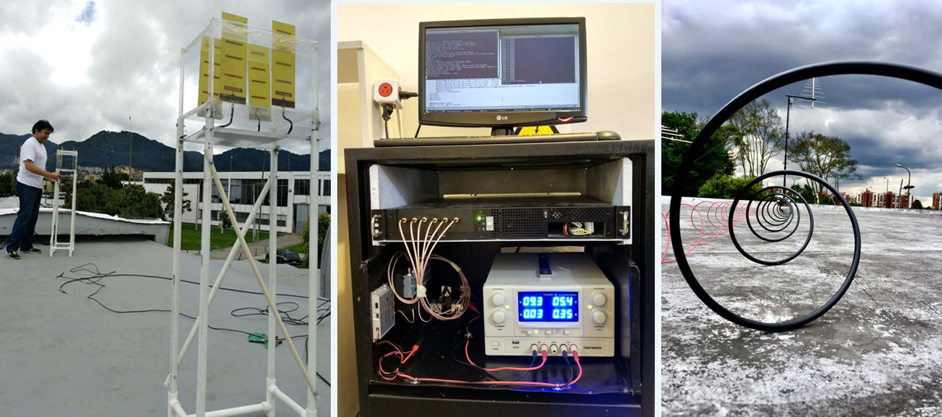}
\par\end{centering}
\caption{\label{fig:Radio-interferometers-currently}Radio interferometers
currently installed on the terrace of the National Astronomical Observatory
at Universidad Nacional de Colombia. The instruments have been
developed by undergrad and master students as part of their research
projects. The First Colombian Radio-interferometer (FiCoRi) (middle and right panels)
represents an important milestone for the development of the radio astronomy
instrumentation in Colombia.}
\end{figure}

In just a few years, the group has shown a rapid evolution and paramount
results evidenced in several scientific publications, master thesis,
undergraduate research works and development of solar instrumentation.
The latter is one of the important milestones in the last couple of
years, with the development of three solar radio interferometers which
are currently installed on the terrace of the building, see Fig.~(\ref{fig:Radio-interferometers-currently}).
These instruments have been fully developed by students in order to
complete their master (2) and undergraduate (1) thesis projects, and
represent the commencement of the instrumental branch at the National
Astronomical Observatory. The research conducted by one of the recently
graduated master student resulted in the development and implementation
of the First Colombian Radio-interferometer (FiCoRi) (right panel
in Fig.~(\ref{fig:Radio-interferometers-currently})) \cite{Vargas2017}. 

Furthermore, and framed in a collaboration with the Planetarium of
Bogot\'a and the District Government, there are plans to implement one
of the interferometers in a large number of public schools, in order
to use big data (astrostatistics) as a tool to promote scientific
knowledge.

Former GoSA members are now pursuing their graduate studies abroad
in high-standard institutes (e.g., University of California Berkeley,
Max Planck Institute for Solar System Research and University of Graz,
among others), and a new generation of local students are willing
to follow up a promising research road, studying the multiple faces
of our active star. 

\section{Discussion and concluding remarks }

The role and importance of astronomy not only for scientific but cultural
development, and a way to answer fundamental questions and driving
innovation, tends to be underestimated, despite of the existence
of a wealth of examples supporting it \cite{Rosenberg2013}. Measuring
scientific development is a difficult task but in general it is related
with the scientific publishing record \cite{Ribeiro2013}, that is
why devoting money and time to all the educational spectrum is paramount.
None of what has been discussed here could be possible without engaging
also administrative and educational heads of the schools and universities,
that provide the instruments and allow to maintain spaces to talk
about science in the classrooms. 

It is important to note explicitly that we did not take an educational theoretical position for all the activities presented here. Instead we have outlined a number of examples from practice to encourage the use of the Sun in science education, informal learning and astronomy communication. Nevertheless, all the experiences took place inside formal educational institutions. We have evidenced that our community has evolved thorough these small and consistent efforts sustained over time, and we have had the experience to see people going through the whole educational spectrum mentioned here, impacting positively our community and country.

Within the general public in Colombia, there is a certain regard of
distrust and even fear of scientific areas. This has been in great
part due to a significant gap existent between scientists and society.
In this sense, science communicators can be fundamental to change
the perspective of the public about science, hence the importance
and need to develop the kind of initiatives mentioned here in the
country, as a way to create spaces that allow the encounter of scientists,
teachers and amateurs to find new and better ways to bring science
to the public. We have experienced how such initiatives definitely
cause a significant impact on students and the way they think about
scientific areas as a future career. 

From the pedagogical point of view, we have learned that it is important
that children and young people recognize the importance of long-term
efforts through continuous observations, which stand as basis to develop
their own research skills working with data. We have also experienced
that students get motivated mainly by visits to astronomical observatories,
attendance to star parties, STEM College and Career Fairs and conferences. Letting scientists
to engage with the public and sharing their personal stories has shown
a unique driven force.

In particular, a project that involves the study of the Sun has shown
in our context, at least, the following advantages:
\begin{itemize}
\item Observations can be made in a regular class time, i.e., around 1.5
hours. 
\item The Sun shows rapid changes that permeates the children with the idea
of the potential to affect the planet in the daily basis. 
\item A Sun-related project can develop associated activities around interdisciplinary
topics, e.g., electromagnetic spectrum, stellar evolution, radio astronomy,
effects of Earth's atmosphere, scientific computing, data analysis
or motors and generators, and ways to create links with biology, chemistry
and even history and archaeology. 
\item The JOVE project offers a great opportunity to participate in scientific
studies with low costs, which typically are under 350 USD, including
the receiver kit, the antenna hardware, the software used and the
tools required for the instrumental assembly.
\end{itemize}
The main goal of the school solar-related experiences that we have
developed in Colombia has been to build up a mechanism to enhance
creativity and to allow participants to acquire and interpret data
in their own ways in order to answer scientific questions. Instructors
are meant to provide an ideal atmosphere to explore ideas, to develop
tools, and help participants with their chosen technique, spreading
their seeds of creativity. 

Studying the Sun has lately shown a prime interest worldwide currently is living a golden age, due to the serious influence of solar
explosive events in our current technological society (in particular
on satellites exposed to the effects of solar activity). At university
level, the field of solar physics has proven to be a source of motivation
for students to pursue further scientific research, mainly because
of all the different aspects that embrace the study of the Sun, from
fundamental physics to data analysis, to develop computational skills for simulation
of solar phenomena and instrumentation. 

As a final remark we want to stress that although solar observation
can be performed relatively easy and at many levels, it is important
to always first prepare students and the general public to avoid the
potential risks that this entails. This should be a major concern
and implies a huge responsibility for the team involved.

{\bf{\textit{Acknowledgements}}}. We would like to thank Juan Camilo L\'opez-Carre\~no and Sebasti\'an Castellanos-Dur\'an, for reading a preliminary version of this manuscript and providing useful feedback.

\end{document}